# Strongly adhesive dry transfer technique for van der Waals heterostructure

*Suhan Son[1,2,3%], Young Jae Shin[4,5%], Kaixuan Zhang[1,2,3], Jeacheol Shin[2], Sungmin Lee[2,3], Hiroshi Idzuchi[4], Matthew J. Coak[2,3,6], Hwangsun Kim[3,7], Jangwon Kim[8], Jae Hoon Kim[8], Miyoung Kim[3,7], Dohun Kim[2], Philip Kim[4], and Je-Geun Park[1,2,3*]*

[1]Center for Quantum Materials, Seoul National University, Seoul, 08826, Republic of Korea

[2]Department of Physics and Astronomy, Seoul National University, Seoul 08826, Republic of Korea

[3]Center for Correlated Electron Systems, Institute for Basic Science, Seoul, 08826, Republic of Korea

[4]Department of Physics, Harvard University, Cambridge, Massachusetts 02138, USA

[5]Center for Functional Nanomaterials, Brookhaven National Laboratory, Upton, New York 11973, USA

[6]Department of Physics, University of Warwick, Gibbet Hill Road, Coventry CV4 7AL, UK

[7]Department of Materials Science and Engineering and Research Institute of Advanced Materials, Seoul National University, Seoul 08826, Republic of Korea

[8]Department of Physics, Yonsei University, Seoul 03722, Republic of Korea

**Abstract.**

That one can stack van der Waals materials with atomically sharp interfaces has provided a new material platform of constructing heterostructures. The technical challenge of mechanical stacking is picking up the exfoliated atomically thin materials after mechanical exfoliation without chemical and mechanical degradation. Chemically inert hexagonal boron nitride (hBN) has been widely used for encapsulating and picking up vdW materials. However, due to the relatively weak adhesion of hBN, assembling vdW heterostructures based on hBN has been limited. We report a new dry transfer technique. We used two vdW semiconductors ($ZnPS_3$ and $CrPS_4$) to pick up and encapsulate layers for vdW heterostructures, which otherwise are known to be hard to fabricate. By combining with optimized polycaprolactone (PCL) providing strong adhesion, we demonstrated various vertical heterostructure devices, including quasi-2D superconducting $NbSe_2$ Josephson junctions with atomically clean interface. The versatility of the PCL-based vdW stacking method provides a new route for assembling complex 2D vdW materials without interfacial degradation.

# 1. Introduction

Mechanical exfoliation and van der Waals stacking of atomically thin materials have been one of the significant experimental methods to produce functional 2D materials heterostructures[1–11]. Experimentally, this technique requires reliable mechanical exfoliation of single crystals on a silicon/$SiO_2$ substrate, followed by picking up or dropping down the atomically thin crystals using various polymer layers such as polymethyl methacrylate (PMMA), polydimethylsiloxane (PDMS), polycarbonyl (PC), and polypropylene carbonate (PPC)[12–15]. However, these polymer layers are known to leave chemical residues on the 2D materials interfaces[12]. Often chemically inert hBN or graphite/graphene layers are used to interface between these polymer layers and target 2D materials for picking up to form vdW heterostructures[9,14–16]. This method has been widely adopted for graphene, hBN, and some other gapped semiconducting transition metal dichalcogenides such as $TMX_2$, where TM=W and Mo, and X=S, Se and Te. However, relatively weak adhesion of hBN and graphene makes it challenging to extend this method to other more complex 2D materials, in particular, metallic and superconducting $TMX_2$ such as TM=Ta and Nb, and X=S, Se and Te.

In this work, we report a novel pick-up method by using optimally formed polymer layers using polycaprolactone (PCL). Furthermore, we developed the technique by combining with vdW large gap semiconductors, i.e., $ZnPS_3$ and $CrPS_4$. Employing improved chemical adhesion of both $ZnPS_3$ and $CrPS_4$ together with mechanical stability provided by PCL, we demonstrate complex vdW heterostructures consisting of superconductors and magnetic materials, where the functional interfaces are completely encapsulated thus free from chemical degradation.

## 2. Method

**2.1. Preparation of PCL stamp.** The PCL stamp was prepared in the following way. First, polycaprolactone (Sigma Aldrich, average Mn: 80,000) was dissolved in tetrahydrofuran (THF) solvent. PCL concentration in the finally obtained solution was 15% in THF (mass percent). Afterward, the mixture was stirred using magnetic stirrer until completely dissolved (> 48 h). The solution formed the viscous and transparent liquid. PDMS was prepared by mixing SYLGARD 184 base and curing agent (10:1 ratio, respectively), followed by curing for 2 hours at 75 ℃. As illustrated in Fig. 1(a), PDMS block was put on top of a glass slide and fixed by the transparent tape (Scotch® Crystal Tape). The PCL solution was then dropped on the tape and spin-coated with the rate of 2,000 rpm for 1 min. and annealed at 75 ℃ for 10 min. to flatten the polymer.

**2.2 Pick-up and drop-down of vdW flakes.** We used a micro-manipulator to align the PCL stamp for the pick-up and drop-down vdW flakes. It was easy to distinguish the area of the PCL stamp attached to the substrate by its color and fringe around it as in ref.[16]. As described in Figure 1(b), the PCL stamp was first put on the substrate at 55 ℃ and then heated to 65 ℃ so that PCL was fully and uniformly melted. Finally, to pick up the flake, the micro-manipulator was retracted very slowly after being cooled to 30 ℃. For the sequential pick-up (Figure 1(d)), the same procedure was repeated. For the drop-down of vdW flakes at the final step, the temperature was again increased to 75 ℃ after touching the substrate. By heating above the melting point of PCL, 60 ℃, we made PCL fully melted and thereby to lose most of the adhesion to the sample, which makes it easy to avoid the unintended pick-up during the drop-down process. Then, the micro-manipulator was slowly retracted, and the PCL stamp was finally isolated to the substrate, leaving all the flakes on the substrate because PCL was still in a liquid state. Every fabrication step was undertaken inside a glove box under an argon atmosphere. To remove the polymer on the flakes, the processed substrate was kept immersed in a THF solvent over-night.

**2.3. Fabrication of $NbSe_2/SiO_2/NbSe_2$ heterostructure.** The $NbSe_2$ flake was exfoliated on 285 nm $SiO_2$ substrate. First, $NbSe_2$ was directly picked up via a PCL stamp, after which $SiO_2$ was deposited on another prepared $NbSe_2$. As the final step, the already picked-up $NbSe_2$ was dropped on $SiO_2/NbSe_2$. After resolving PCL by THF, the electrode was deposited via a conventional e-beam lithography technique and e-beam evaporation. The metal electrode of 5 nm Ti and 60 nm Au were deposited on top of samples by the electron-beam evaporator.

**2.4. Fabrication of $CrPS_4/NbSe_2/CrPS_4$ heterostructure.** At first, all samples were exfoliated on 285 nm $SiO_2$ substrate. Then, the PCL stamp was used to pick up the top-most $CrPS_4$, which was followed by picking up $NbSe_2$ and $CrPS_4$ sequentially. Moreover, the combined structure of $PCL/CrPS_4/NbSe_2/CrPS_4$ was dropped on the pre-patterned electrodes that were prepared in the same way as that of the NbSe$_2$/SiO$_2$/NbSe$_2$ heterostructure. Metal contacts of 2 nm Ti and 8 nm Pt were deposited on top of samples by the electron-beam evaporator.

**2.5. Cs-TEM measurement**. Samples were prepared with a focused ion beam (FIB) milling with a FIB instrument (Helios 650, FEI) and ion milling with a nanomill (Model 1040, Fischione). Atomic resolution high angle annular dark-field scanning transmission electron microscopy (TEM) image and energy dispersive spectroscopy results were obtained with 80 kV spherical aberration-corrected TEM (JEM-ARM200F, JEOL).

**3. Results and discussion**

Layered transition metal thiophosphate (*TM*PS$_x$) is a class of 2D vdW semiconductors that exhibit various magnetic properties[17,18]. For example, ZnPS$_3$ is a paramagnetic semiconductor, while CrPS$_4$ is an antiferromagnetic insulator with a Neel temperature of 36 K[19]. With the robust adhesive strength of PCL, we improved the chance of picking up van der Waals flakes. We also showed that with a PCL stamp, an insulating vdW ZnPS$_3$ could be used to pick up other vdW materials, which were previously known to be very challenging while maintaining a clean interface (Figure 1(e)-(h) & 2(b)).

To fabricate novel vdW heterostructures, we have developed the PCL-based stamp to handle vdW flakes, which has been technically challenging using the conventional polymer/hBN based method. The schematics of the PCL stamp are illustrated in Figure 1(a) and (b). By adhesive strength measurements, we found that PCL has the most robust adhesive strength than any other polymers known until now, at least 20 times larger than that of the most commonly used PPC[14] (see Supplementary Note 1). As a real test of the PCL stamp, we could successfully pick up all exfoliated graphite on oxygen plasma treated $SiO_2$ substrate and $CrPS_4$ (see Supplementary Note 1).

In the conventional vdW stacking method, the PPC or PC-based hBN pick-up method has been widely used as a top-most layer for the vdW heterostructure. Chemically inert hBN exhibits weak vdW interaction between hBN and substrate. This weak adhesion allows picking up hBN flakes from the $SiO_2$ substrate, where the flakes are initially deposited during the mechanical exfoliation process[14]. The PPC or PC polymer layers provide a rather weak adhesive strength, just enough to pick up hBN, making other vdW materials for the initial picking up layer for vdW heterostructure sub-optimal.

Once hBN is picked up, one can use the vdW force between hBN and the target flake to fabricate heterostructures. In this way, the PPC/hBN pick-up method has been able to cover many parts of the finally built vdW heterostructures. However, hBN is known to be bad at detaching some vdW materials from the substrate, making the PPC/hBN pick-up method more challenging and time-consuming for certain vdW materials[20]. To successfully pick up vdW flakes, the adhesive vdW force between the top-most material and the target flake must be significantly stronger than that between the substrate and the target flake. However, if the target flakes have strong adhesion with the deposited substrate, the PPC/hBN pick-up method fails. In particular, $NbSe_2$, a 2D vdW superconductor, is known extremely hard to pick up. Several innovative experimental techniques, including the Via method, electrical contact with graphite, and PDMS drop down[20–23], have been performed to build vdW devices together with $NbSe_2$. Still, very limited device fabrication yields have so far been reported due to strong adhesion of NbSe$_2$ to the SiO$_2$ substrates. We found that the strong adhesive force of PCL provides enough adhesion to pick up several new vdW flakes directly from the substrate. Therefore, there is no need to restrict the top-most vdW layer to hBN in the PCL-based pick-up method.

The strong adhesion using PCL allows us to apply this polymer layer to pick up other designed vdW materials as the top-most layer for vdW heterostructures. Using this method, we first tested the insulating vdW material, ZnPS$_3$ (see Supplementary Note 2), as the top-most layer. The PCL stamp was used to pick up ZnPS$_3$ from the substrate in Figure 1(c).

We also found that the PCL/ZnPS$_3$ can pick up the other target vdW flakes with high yield (Figure 1(d)). Figure 1(e)-(g) showcase the successful examples of several vdW target flakes, which we tried with this method with the following success rate: 12 success/14 trials, 2/2, 2/3, and 1/2 for NbSe$_2$, Fe$_3$GeTe$_2$, NiPS$_3$, and VSe$_2$, respectively. In contrast, when we used the PPC/hBN pick-up method, it produced significantly smaller yields: 0 success/7 trials, 0/4, 1/4, and 0/1 for NbSe$_2$, Fe$_3$GeTe$_2$, NiPS$_3$, and VSe$_2$, respectively. With this new method reported here, we succeeded in all the van der Waals materials we tested: NbSe$_2$, Fe$_3$GeTe$_2$, NiPS$_3$, and VSe$_2$. Furthermore, we also found in this study that ZnPS$_3$ is a right candidate as an alternative top-most capping layer, aspects of which provides the adhesion force enough to target material under fabrication and the material as the protecting layer for underlying flakes (see Supplementary Note 3). Using PCL, we also demonstrated a direct pick-up method successfully without any top-most capping layer, as shown in Figure 2(a) and 3(a). In this case, the functionalized van der Waals materials were used as the self-defined top-most capping layer.

A clean interface also demonstrates the strength of this new pick-up method. Among the all vdW heterostructure assembly methods, the vdW pick-up method by hBN shows the cleanest interface[12]. It is because vdW flakes are sequentially picked up, which makes their interfaces, not in direct contact with the polymer. Therefore, underneath flakes and its interfaces remain free of the polymer residue. We demonstrate that it is also the case with the PCL pick-up method. We checked the interface in the test samples of $NbSe_2/SiO_2/NbSe_2$ and $ZnPS_3/NbSe_2$ heterostructures (see Figure 2) with the transmission electron microscope (TEM) images (insets of Figure 2(a) and (b)) showing a very sharp and clean interface.

Moreover, there is no artificial effect in the transport data originating from the fabrication method (see Supplementary Note 3). All these observations validate the clean interface achieved by the PCL pick-up method. A clean interface between two different materials is crucial to attaining new emergent phenomena[24–27]. Thanks to its clean interfaces by the pick-up method, we can now explore the intriguing phenomena that could otherwise be more difficult, if not impossible, to realize and thereby investigate novel physics like proximity effect at the interface of the vdW heterostructure system.

As another test, we fabricated the $CrPS_4/NbSe_2/CrPS_4$ heterostructure (Figure 3(a) and (b)). It not only introduces the magnetic proximity effect to the superconducting system in vdW heterostructure[28] but also demonstrates the novel aspect of the PCL method further. Here we could expect the additional magnetic exchange interaction to $NbSe_2$ due to the nearby A-type antiferromagnetic vdW $CrPS_4$, which is different from the case of non-magnetic insulator $ZnPS_3$ described in Figure 2(b) (see Supplementary Note 2). And the NbSe$_2$ superconductivity is expected to be affected dramatically by the magnetic moments of CrPS$_4$[26]. We note that because the clean interface is crucial and ferromagnetic insulators are quite rare, this kind of experiment has remained limited only to a handful of cases[29].

For the fabrication of the heterostructure (Figure 3(a) and (b)), we sequentially picked up the three key materials from the SiO$_2$ substrates: the top CrPS$_4$, the middle NbSe$_2$, and the bottom CrPS$_4$ (see Methods). Note that we were able to pick up NbSe$_2$ using CrPS$_4$. We emphasize that during the whole process, the middle NbSe$_2$ layer was never exposed to the PCL polymer, unlike the previously known methods[12].

The quality of CrPS$_4$/NbSe$_2$ interfaces can further be tested by carrying out electrical transport measurements, as shown in Figure 3(b). In this heterostructure, the electrical current flows only through NbSe$_2$ since CrPS$_4$ is insulating[30]. All the measurements were performed at 1.5 K, well below the T$_C$ of NbSe$_2$. Figure 3(c) shows the typical superconducting behavior of NbSe$_2$ across the region, where a single side of NbSe$_2$ is in contact with CrPS$_4$. Interestingly, we find that the superconducting behavior of NbSe$_2$ destroyed across the part where both sides of NbSe$_2$ are sandwiched by CrPS$_4$ (Figure 3(d)), exhibiting metallic V-I characteristics (non-zero slope near zero-current). Assuming the quality of NbSe$_2$ is similar in both regions, the observation of non-superconducting behavior in the CrPS$_4$-sandwiched part of the NbSe$_2$ sample suggests that superconductivity of NbSe$_2$ was significantly weakened by the spin alignment nearby from the magnetic material. Similar destruction of superconductivity was reported in the thin films of ferromagnet and superconductor[31].

## 4. Conclusions

In conclusion, our new PCL-based pick-up method, together with $ZnPS_3$ and $CrPS_4$, can be used for several vdW flakes that have been known difficult to handle thanks to its strong adhesive strength. By using this method, several new vdW heterostructures, otherwise challenging to fabricate, could be achieved with a unique combination and an increased yield. Furthermore, we could also show that this original pick-up method ensures a polymer-residue-free and clean interface by TEM measurement and transport measurements. Finally, we succeeded in fabricating $NbSe_2$ heterostructure sandwiched by magnetic $CrPS_4$, showing an apparent magnetic proximity effect. The PCL pick-up method presented here will be proven invaluable to the much more extensive efforts towards the fabrication of clean-interfaced vdW heterostructure boosting the related research field.

**Figures**

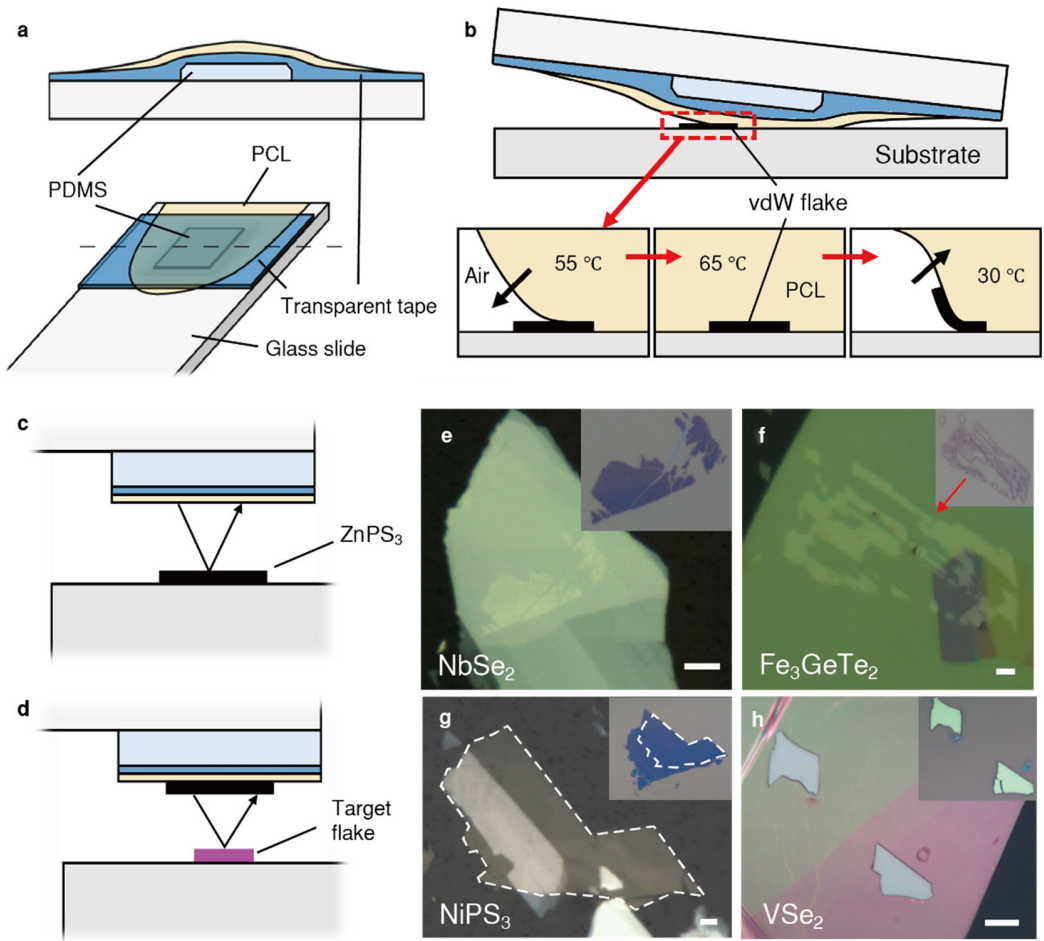

**Figure 1.** Schematics of dry transfer technique and its pick-up ability. (a) The structure of the PCL stamp for the fabrication. (b) Schematics of the pick-up process. (c) ZnPS$_3$ picked-up by PCL stamp. (d) Continuous pick-up of target flake followed by c (with ZnPS$_3$, here). (e)-(h) Optical images of PCL stamp after the pick-up process. Examples of NbSe$_2$, Fe$_3$GeTe$_2$, NiPS$_3$, and VSe$_2$ were shown, respectively. A bright color difference can recognize the shape of target flakes in each inset compared to the background of the ZnPS$_3$ flake in the image. All scale bars are for 5 μm. Insets: optical images of exfoliated target flakes on 285 nm SiO$_2$ substrate before the pick-up process.

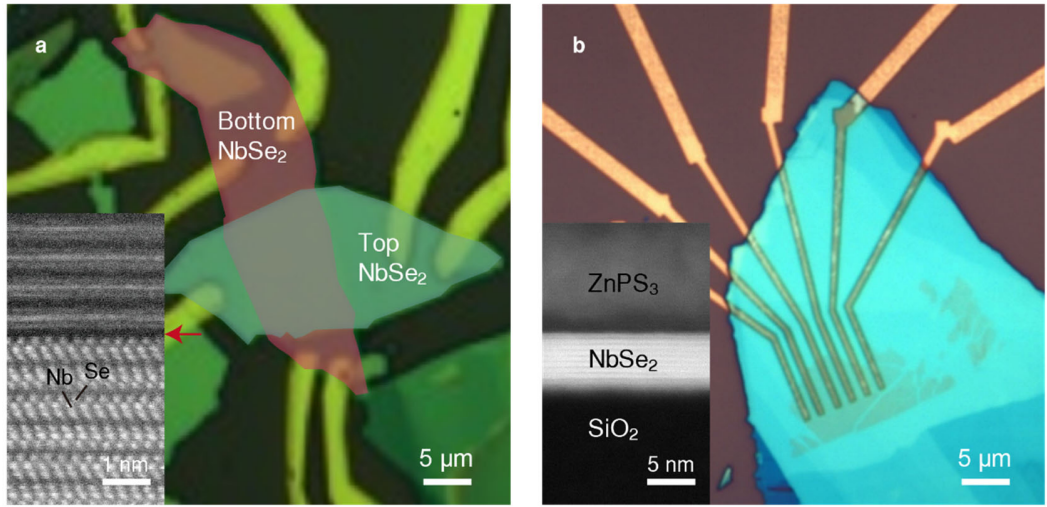

**Figure 2.** NbSe$_2$ heterostructures and its TEM images. (a) Optical microscope image of NbSe$_2$/SiO$_2$/NbSe$_2$ Josephson junction. Top NbSe$_2$ was directly detached from the SiO$_2$ substrate using a PCL stamp. Top and bottom NbSe$_2$ are depicted as the false-color for clarity. Inserted SiO$_2$ layer was 0.13 nm thick. (b) Optical microscope image of ZnPS$_3$/NbSe$_2$ flakes. It was dropped down on the pre-patterned electrode on 285 nm SiO$_2$ substrate after being picked up sequentially from the substrate. Insets: TEM images of each device showing a clean and sharp interface of the heterostructure.

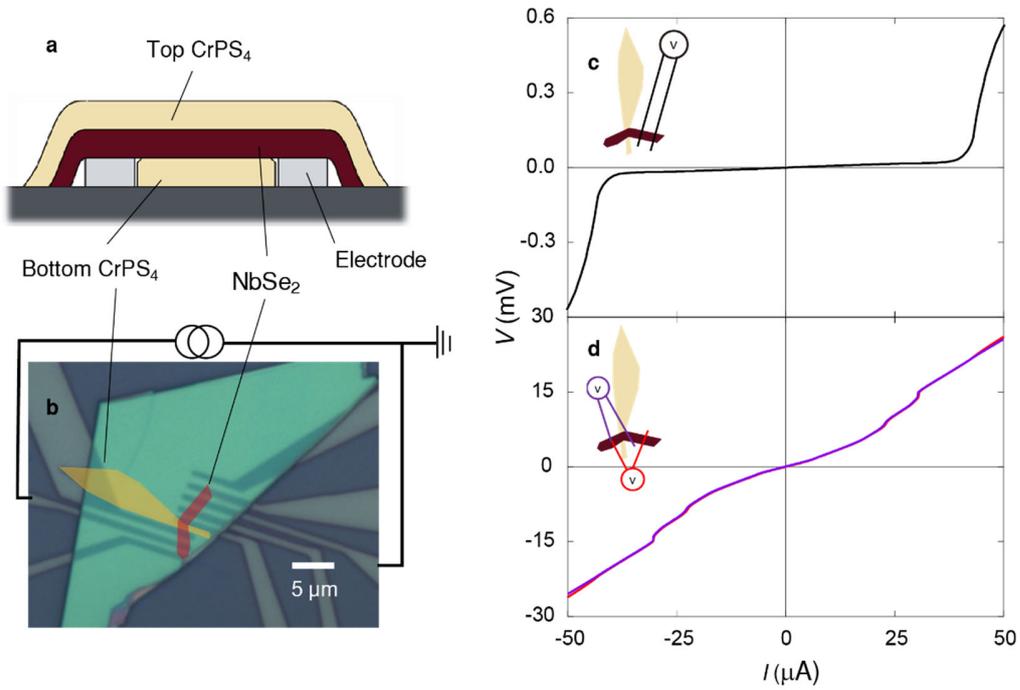

**Figure 3.** Transport measurement of $CrPS_4/NbSe_2/CrPS_4$ heterostructure. (a) Schematics of the device. (b) Optical microscope image of the device (False color). Bottom $CrPS_4$ and $NbSe_2$ are colored to the same color to the corresponding layer of a for clarity. (c) and (d) V-I characteristics at 1.5 K. Voltage drop was measured at different parts. Each colored data indicates the same colored wire configuration drawn at the schematics in the inset.

**Acknowledgments**

The authors would like to thank Jonghyeon Kim, Inho Hwang, Youjin Lee, and Junghyun Kim for their generous help and discussion. Work at CQM was supported by the National Research Foundation of Korea (Grant No. 2020R1A3B2079375). Work at IBS CCES was supported by the Institute for Basic Science of the Republic of Korea (Grant No. IBS-R009-G1); work at Harvard Univ. was supported by Global Research Laboratory Program (2015K1A1A2033332) through the National Research Foundation of Korea (NRF) and DOE QPRESS; work by Y.J.S. used resources at the Center for Functional Nanomaterials, which is a U.S. DOE Office of Science Facility, and at Brookhaven National Laboratory under Contract No. DE-SC0012704; work at Yonsei Univ. was supported by the National Research Foundation of Korea (NRF) grants funded by the Ministry of Science, ICT (SRC Program No. NRF-2017R1A5A1014862); work of M.K. was supported by the National Research Foundation of Korea (NRF) grants funded by the Ministry of Science and ICT (MSIT) [No. 2015R1A5A1037627]; work of D.K. was supported by Basic Science Research Program, Quantum Computing Technology Development Program (Grant No. 2018R1A2A3075438 and No. 2019M3E4A1080145) through the National Research Foundation of Korea (NRF) funded by the Korean government (MSIT) and Creative-Pioneering Researchers Program through Seoul National University (SNU).

**Author contributions**

SS and YJS contributed equally. J-GP and PK supervised the project. SS, YJS, and SL prepared devices and single crystals, and YJS developed the PCL polymer. SS, JS, HI, and DK measured transport properties. HK & MK performed Cs-TEM measurement, and JK & JHK carried out optical absorption measurement. SS, YJS, KZ, MJC, PK, and J-GP designed the experiments, and SS and YJS analyzed the data. All authors discussed the results, and SS, KZ, PK, and JGP wrote the manuscript with comments from all authors.

**Additional Information**.

Supplementary information is available in the online version of the paper.

**Competing financial interests**

The authors declare no competing financial interest.

# Supplementary information for

# Strongly adhesive dry transfer technique for van der Waals heterostructure

Suhan Son[1,2,3%], Young Jae Shin[4,5%], Kaixuan Zhang[1,2,3], Jeacheol Shin[2], Sungmin Lee[2,3], Hiroshi Idzuchi[4], Matthew J. Coak[2,3,6], Hwangsun Kim[3,7], Jangwon Kim[8], Jae Hoon Kim[8], Miyoung Kim[3,7], Dohun Kim[2], Philip Kim[4], and Je-Geun Park[1,2,3]*

[1]Center for Quantum Materials, Seoul National University, Seoul, 08826, Republic of Korea

[2]Department of Physics and Astronomy, Seoul National University, Seoul 08826, Republic of Korea

[3]Center for Correlated Electron Systems, Institute for Basic Science, Seoul, 08826, Republic of Korea

[4]Department of Physics, Harvard University, Cambridge, Massachusetts, 02138, USA

[5]Center for Functional Nanomaterials, Brookhaven National Laboratory, Upton, New York 11973, USA

[6]Department of Physics, University of Warwick, Gibbet Hill Road, Coventry CV4 7AL, UK

[7]Department of Materials Science and Engineering and Research Institute of Advanced Materials, Seoul National University, Seoul 08826, Republic of Korea

[8]Department of Physics, Yonsei University, Seoul 03722, Republic of Korea

**%** Authors with equal contributions

* Corresponding authors. e-mail: jgpark10@snu.ac.kr

Contents:

Supplementary Note 1. Comparison of the physical properties and pick-up ability of polymers including our new polycaprolactone (PCL)

Supplementary Note 2. Crystal growth and elementary characterization of bulk crystals: $ZnPS_3$, $NbSe_2$, and $CrPS_4$

Supplementary Note 3. Transport data of the heterostructure shown in Figure 2 and the ability of $ZnPS_3$ as the capping layer

Supplementary Note 4. Additional transport data for Figure 3

Figure S1. Physical properties of polymers and comparison to the PPC dry transfer method

Figure S2. Basic characterization of $ZnPS_3$

Figure S3. Magnetic properties of $CrPS_4$

Figure S4. Transport data of $NbSe_2$/$NbSe_2$ heterostructure device described in Figure 2a

Figure S5. Transport data of $ZnPS_3$/$NbSe_2$ heterostructure device described in Figure 2b

Figure S6. Additional transport data for the $ZnPS_3$/$NbSe_2$ heterostructure

Figure S7. Cs-corrected transmission electron microscope (Cs-TEM) measurement of $ZnPS_3$/$NbSe_2$ and energy dispersive X-ray spectroscopy (EDX)

## Supplementary Note 1. Comparison of the physical properties and pick-up ability of polymers including our new polycaprolactone (PCL)

**Adhesive strength measurement.** Adhesive measurement was conducted using a universal test machine (Lloyd LR 10K) to detect the strength of the adhesion (Figure S1a). PCL used in this study has a molecular weight of 80,000, showing ~ 5 MPa of adhesive strength. For comparison, we also measured the adhesive strength of polycarbonate (PC) and polypropylene carbonate (PPC), which are frequently used polymers for the fabrication of van der Waals (vdW) heterostructure[1]. Among them, the adhesive strength of PCL showed the highest value. Each chemical structure of polymers was also illustrated in Figure S1a.

**DSC measurement.** Differential scanning calorimetry (DSC) studies were conducted using a differential scanning calorimeter (Scinco DSC N-650) under a nitrogen atmosphere to detect the thermal behavior of the polymeric materials (Figure S1b). PCL shows the melting temperature ($T_m$) of 60 °C. The glass transition temperature ($T_g$) of PCL, PPC, and PC was determined to be -60, 45 and 145 °C, respectively.

**Pick-up ability test.** To examine the pick-up ability using real materials, we prepared graphites exfoliated on oxygen plasma treated 285 nm $SiO_2$ substrate. The flow rate of oxygen gas for plasma treatment on substrates was kept at 100 sccm, and the acceleration power was set to 300 W. And the pressure of the chamber was monitored at 300 mTorr. After 10 minutes of plasma treatment, graphite was exfoliated. This process is known to increase the attractive force between vdW flakes and the substrate[2]. The prepared graphites are shown in Figs. S1c and S1e. Both PCL and PPC stamps were used to pick up these graphites. The PCL stamp succeeded in picking up every graphite flake in the image, while the PPC stamp could pick up only thick (yellowish) ones. Figs S1d and S1f show the optical images of the PCL and PPC stamps after pick-up. The red arrows in Figure S1e indicate the graphites remaining on the substrates after pick-up. The pick-up yield was 100% (15/15 flakes) for the PCL stamp and 68.4% (13/19 flakes) for the PPC stamp. However, the pick-up probability of the PPC stamp decreased abruptly in thin graphite layers (Bluish flakes in figure S1e).

Furthermore, $CrPS_4$ was also tested by the same method (Figs. 1g-1j). The yield of the pick-up was close to 98.1% (101/103 flakes) for PCL stamp, while only 25.5% of flakes (12/47) were able to be picked by PPC stamp solidly. Among them, thin flakes (< 40 nm) show a very low probability, like graphite (white and blue flakes in figure S1i). These results indicate that PCL provides a more efficient way to pick up van der Waals flakes with the increased adhesion force. Therefore, this method can be a useful tool for versatile applications: for example, fabricating vdW heterostructure or twisted vdW heterostructure, or transferring flakes to other substrates.

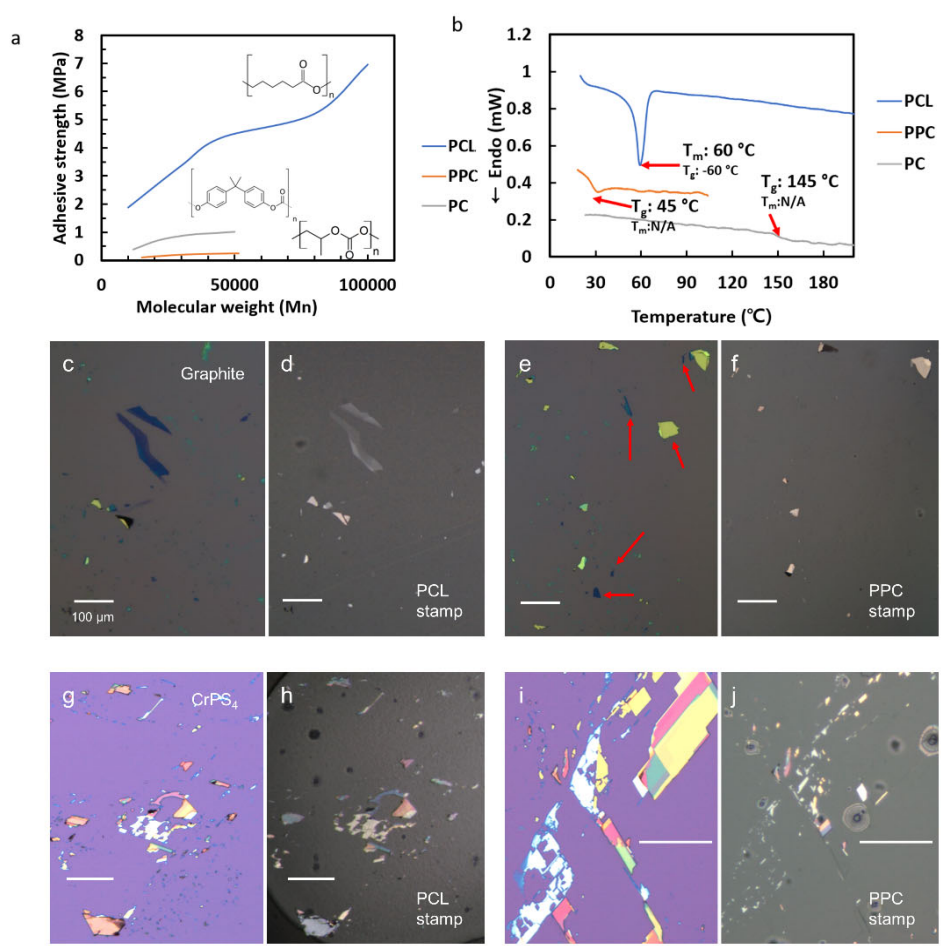

**Figure S1.** Physical properties of polymers and comparison of the pick-up ability of PPC and PCL. (a) The adhesive strength of polypropylene carbonate (PPC), polycarbonate (PC), and polycaprolactone (PCL) depending on molecular weight. (b) DSC data of each polymer. (c) Graphite exfoliated on oxygen plasma treated $SiO_2$ substrate. (d) Optical microscope (OM) image of the PCL stamp after the pick-up of graphites shown at (c). Picked-up graphites are distinguishable as a bright gray color. Every flake on $SiO_2$ could be picked up from the substrate. (e) same as (c). (f) OM image of the picked-up graphite flakes by the PPC stamp. Flakes remaining on $SiO_2$ substrate were marked as red arrows on (e). (g), (i) OM image of exfoliated $CrPS_4$ on $SiO_2$ substrate. (h), (j) OM image of picked-up $CrPS_4$ via PCL stamp and PPC stamp, respectively. All scale bars represent 100 μm.

# Supplementary Note 2. Crystal growth and elementary characterization of bulk crystals: $ZnPS_3$, $NbSe_2$, and $CrPS_4$

Single crystals of $ZnPS_3$, $NbSe_2$, and $CrPS_4$ were all synthesized via a self-flux chemical vapor transport (CVT) as the ref.[3].

**Synthesis of single crystals $ZnPS_3$.** For $ZnPS_3$, we mixed zinc metallic powder (Sigma-Aldrich, >99.995%), phosphorous red (Sigma-Aldrich, >99.99%) and sulfur flakes (Sigma-Aldrich, >99.998%) in the stoichiometric ratio with additional 5% of sulfur for the transport agent inside the Ar-filled glove box and sealed. The sealed quartz tube was placed in the two-zone furnace to initiate the growth. The furnace was heated to 540/470 °C for one week and slowly cooled down to room temperature. The final crystals formed in shiny transparent plates. $ZnPS_3$ is stable in the air, and there is no sign of degradation even after a few weeks of exposure to the air.

**Basic characterization of $ZnPS_3$.** The stoichiometric ratio of bulk $ZnPS_3$ crystal was further analyzed by an energy dispersive X-ray (EDX) spectroscopy (not shown). For the bandgap analysis, the optical absorption spectroscopy measurement was carried out (Figure S2a). The measured indirect bandgap was determined to be 3.32 eV by transmittance measurements and its Tauc plot (inset of Figure S2a), similar as reported by Du *et al*.[4]. And this value is the second-largest bandgap among van der Waals crystals with the largest value of 6 eV published for hBN[5]. X-ray diffraction was further performed to check the crystal quality. The simulated (00l) Bragg peaks are well matched to the experimental results (Figure S2b).

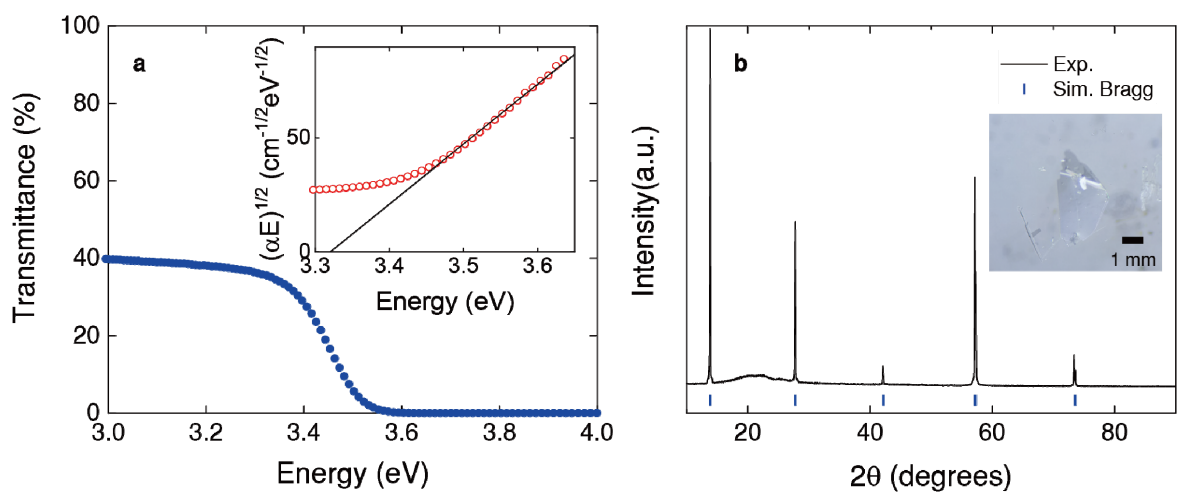

**Figure S2.** Basic characterization of $ZnPS_3$. (a) The transmittance measurement result of $ZnPS_3$. Measured indirect bandgap from the absorption coefficient is 3.32 eV (inset) (b) X-ray diffraction and simulated Bragg peaks. Experiment and simulation results are well-matched with one another. Inset: Image of the single crystal.

**Synthesis of single crystals $NbSe_2$.** For $NbSe_2$, we mixed niobium metallic powder (Alfa Aesar, >99.99%) and selenium powder (Sigma-Aldrich, >99.99%) inside the Ar-filled glove box in the stoichiometric ratio with additional 1% of iodine for the agent and sealed. The sealed quartz tube was placed in the two-zone furnace to initiate the growth. The furnace was heated to 730/700 °C for 3 weeks and slowly cooled down to room temperature. The final crystals formed in shiny and black plates. The stoichiometric ratio was further checked by an energy dispersive X-ray (EDX) spectroscopy, and SQUID-VSM confirmed its bulk superconductivity with perfect diamagnetism. And it shows zero resistance, which is verified by transport measurement below $T_C$ (not shown).

**Synthesis of single crystals CrPS$_4$.** For CrPS$_4$, we mixed chromium metallic powder (Alfa Aesar, >99.996%), phosphorous red (Sigma-Aldrich, >99.99%) and sulfur flakes (Sigma-Aldrich, >99.998%) inside the Ar-filled glove box in the stoichiometric ratio with additional 5% of sulfur for the agent and sealed as ZnPS$_3$. The sealed quartz tube was placed in the two-zone furnace to initiate the growth. The furnace was heated to 650/550 °C for one week and slowly cooled down to room temperature. The final crystals formed in shiny dark reddish plates with one axis of single-crystal longer than the others. The stoichiometric ratio was further checked by an energy dispersive X-ray (EDX) spectroscopy (not shown). CrPS$_4$ is so stable in the air that we cannot note any sign of degradation during the experiment.

**Magnetism of CrPS$_4$.** Our susceptibility measurement taken by SQUID-VSM (Figs. S3a and b) shows that the magnetic ground state is antiferromagnetic with the easy axis of the out-of-plane axis (c* axis). We also obtained from the same data that the Curie-Weiss temperature is 41 K. Note that this large positive value of the Curie-Weiss temperature for an antiferromagnet implies that it is an A-type antiferromagnet with an in-plane ferromagnetic arrangement like FeCl$_2$[6]. This conclusion is consistent with the DFT result[7].

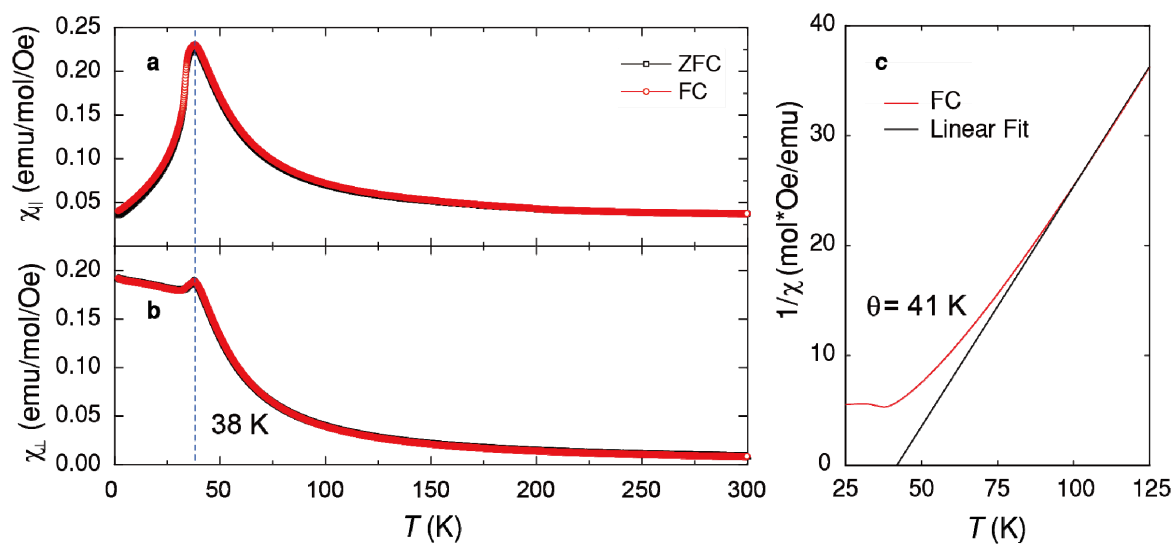

**Figure S3.** Magnetic properties of CrPS$_4$. (a) Susceptibility (ZFC: zero-field-cooled and FC: field-cooled) measured along the c* axis and (b) perpendicular to the c* axis. Neel temperature, $T_N$, was determined to be 38 K. (c) Inverse magnetic susceptibility of (b). Measurement was performed in the in-plane magnetic field with the Curie-Weiss temperature of 41 K.

## Supplementary Note 3. Transport data of the heterostructure shown in Figure 2 and the ability of ZnPS$_3$ as the capping layer

**Josephson junction behavior of NbSe$_2$/SiO$_2$/NbSe$_2$ shown in Figure 2a.** The transport measurement on the NbSe$_2$/SiO$_2$/NbSe$_2$ Josephson junction was carried out. Here, inserted SiO$_2$ (0.13 nm thick) was introduced as a weak link. Furthermore, the discrepancy of the aligned angle of each NbSe$_2$ layer was distinguishable at the vdW interface in the TEM image (inset of Figure 2a), which also acts as a weak link of the Josephson junction. V-I characteristics (Figure S4a) at different temperatures were measured, showing a qualitatively same behavior to the Josephson junction shown in the ref.[8]. Additionally, from the temperature-dependent critical currents, the superconducting energy gap at 0 K can be extracted by using both Ambegaokar-Baratoff (AB) and Bardeen-Cooper-Schrieffer (BCS) theories. The AB theory can be expressed as

$$\frac{I_c(T)}{I_c(0)} = \frac{\Delta(T)}{\Delta(0)} \tanh[\frac{\Delta(T)}{2k_B T}]$$

, where $I_c(0)$ is the zero-temperature critical current, $\Delta(0)$ is the zero-temperature superconducting gap and $k_B$ is the Boltzmann constant [9]. And by the BCS theory, $\Delta(T)$ can be expressed as

$$\Delta(T) = \Delta(0)\tanh(2.2\sqrt{(T_c - T)/T})$$

, where $T_c$ is the critical temperature. From these relations, $\Delta(0)$ was extracted to be 0.7 meV, which is consistent with the previous report[8].

Furthermore, the Josephson critical current density was determined to be $0.036 \mu A/\mu m^2$ for the 0.13 nm SiO$_2$ case. We note that the Josephson current density is significantly suppressed by 3 orders of magnitude as compared with the case of thicker SiO$_2$ barrier (0.4 nm SiO$_2$ barrier, not shown). It demonstrates that the Josephson coupling was decreased as the thickness of the insulating barrier was increased. All these observations are well matched to the conventional effect of Josephson junction.

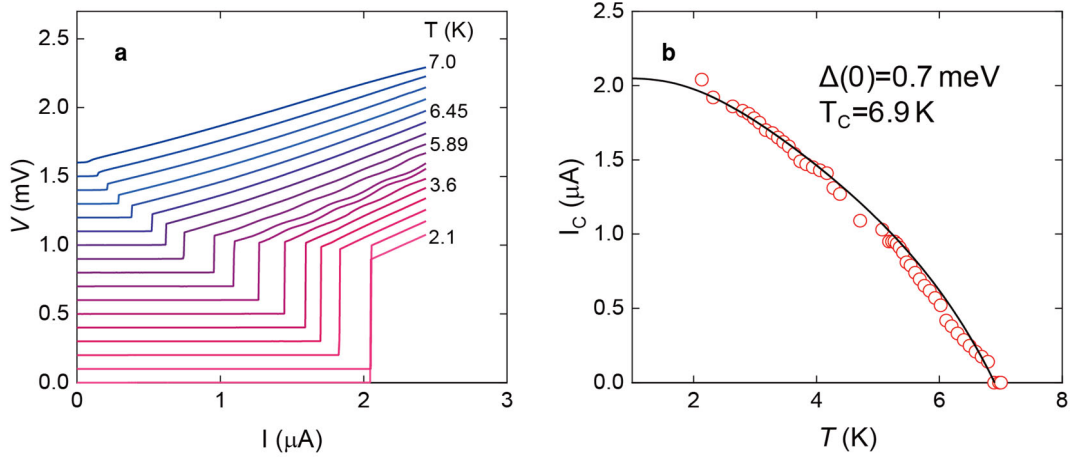

**Figure S4.** Transport data of NbSe$_2$/NbSe$_2$ heterostructure device described in Figure 2a. (a) Temperature-dependent V-I characteristics. For a better presentation, each data were shifted vertically. (b) Temperature-dependent critical current. The AB theory fits the black line.

**Transport data of ZnPS$_3$/NbSe$_2$ heterostructure shown in Figure 2b and capping the ability of ZnPS$_3$.** Because ZnPS$_3$ has a large bandgap (3.32 eV) and is an excellent insulator, it does not affect the transport measurement at all. Figure S5 shows the transport result of the ZnPS$_3$/NbSe$_2$ heterostructure shown in Figure 2b. Fig S5a shows the temperature-dependent resistance with superconducting behavior below T$_C$. The device was fabricated inside an Ar-filled glove box. However, it was intentionally exposed to the air for 72 h after the first measurement to show the capping ability of ZnPS$_3$ with no visible sing of changes in T$_C$ after 72 h. Figures S5b and S5c show the temperature and field dependency of resistance with varying perpendicular magnetic fields and temperatures. All data exhibit the typical superconducting behavior of NbSe$_2$[10].

The capping ability of ZnPS$_3$ was further examined by another configuration. Now, ZnPS$_3$ did not fully cover NbSe$_2$, and a small portion of NbSe$_2$ was exposed to the air. Then, the whole layers were picked up and dropped down on the pre-patterned electrodes, as shown in Fig S6a. Initially, both layers were the same flake having the same thickness. The transport results taken on the two parts of the sample, however, show the different results, as shown in Fig S6b. The exposed part of NbSe$_2$ has a lower T$_C$ than the covered part, possibly due to several factors: one is the contamination from polymer residue and another a different level of degradation from the exposure to the outside atmosphere. Field and temperature dependency were also measured in the capped part (Figure S6c and S6d), like Figure S5. Here, all transport measurements were carried out using a Teslatron PT (Oxford instruments) and by a typical lock-in technique. The current was set to 500 nA, small enough to minimize the effect of current. Therefore, we can conclude from the three different transport measurements that ZnPS$_3$ acts well as a capping layer, and the pick-up method is working, as mentioned in the main text.

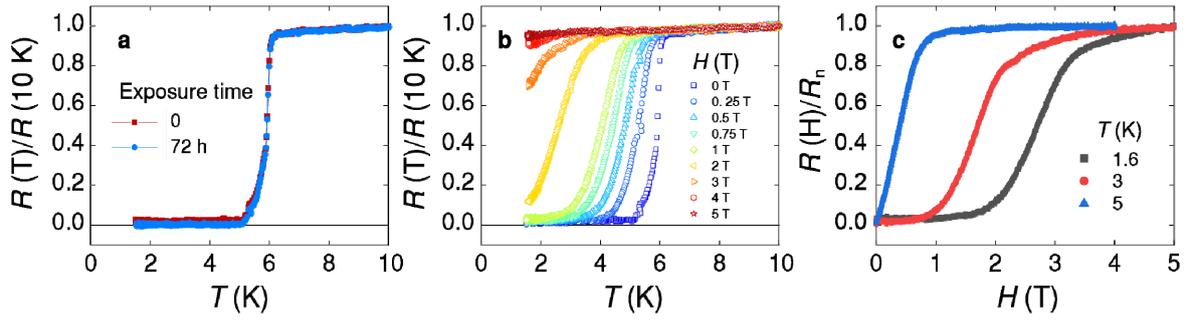

**Figure S5.** Transport data of ZnPS$_3$/NbSe$_2$ heterostructure device described in Figure 2b. (a) Temperature dependence of the resistance for the device. During the device fabrication process, the whole flakes were kept under the Ar atmosphere. No change of T$_C$ was observed after the intentional exposure in the air for 3 days. (b) R(T) measured under the out-of-plane magnetic field. (c) Magnetic field dependence of resistance at different temperatures. All data show the superconducting behavior of NbSe$_2$ itself.

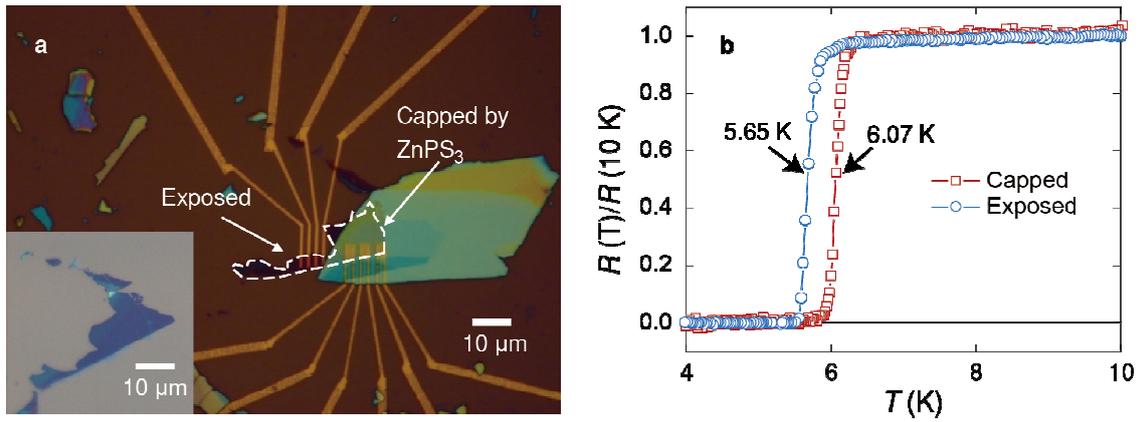

**Figure S6.** Additional transport data for the ZnPS$_3$/NbSe$_2$ heterostructure. (a) The photo of the sample prepared for the measurement. Exposed and capped NbSe$_2$ samples have the same thickness. Inset, NbSe$_2$ flakes on 285 nm SiO$_2$ substrate before pick-up. (b) Temperature-dependent resistance was measured at two different parts: The part capped by ZnPS$_3$ (Capped) and exposed to the air (Exposed). (c) R(T) measured under the out-of-plane magnetic field and (d) magnetic field-dependence of resistance at different temperatures of the exposed part. This result indicates that ZnPS$_3$ can protect the sample from possible chemical attacks from the outside atmosphere or possible contamination coming from polymer residue, and ZnPS$_3$ itself is transparent.

**EDX line scan of ZnPS$_3$/NbSe$_2$ heterostructure.** EDX line scan was carried out on the sample shown in Fig 2b (Figure S7).

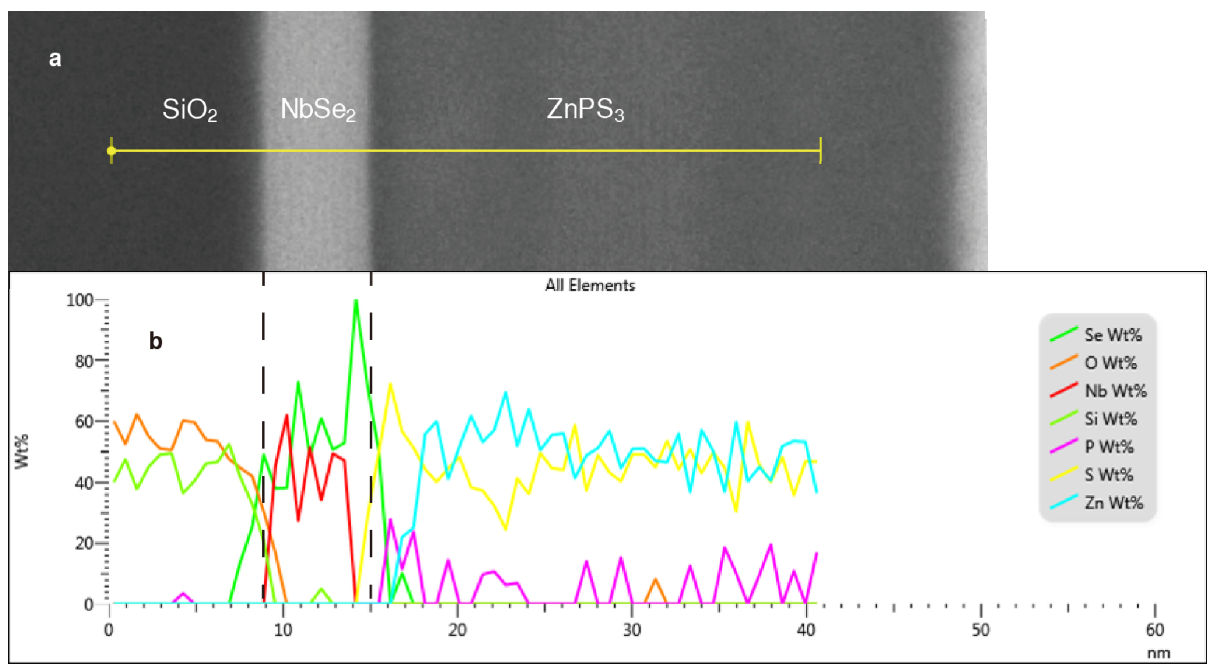

**Figure S7.** Cs-corrected transmission electron microscope (Cs-TEM) measurement of ZnPS$_3$/NbSe$_2$ and energy dispersive X-ray spectroscopy (EDX). (a) Cross-sectional view. (b) EDX line scan marked on (a). A clear boundary was shown by color difference. All compounds containing heterostructure were matched in tendency.

# Supplementary Note 4. Additional transport data for Figure 3

We carried out all the measurements shown in Figs. 3a and 3b for the CrPS$_4$/NbSe$_2$/CrPS$_4$ heterostructure in the following manner: temperature dependency, V-I characteristics, and direct dV/dI measurement of the heterostructure. The voltage drop was measured across the region marked as the red line in Figure 3c (across the area where two sides of NbSe$_2$ contacted CrPS$_4$). The temperature-dependent resistance was measured by a lock-in technique (Figure S7a) with the RMS current of 500 nA. It shows a sudden voltage drop at the temperature corresponding to T$_C$ of NbSe$_2$ (Black arrow in Figure S8a) but did not reach to zero resistance. For the direct dV/dI measurement, we applied DC current and an additional small amount of oscillating current (< 5%) at each DC current level and simultaneously measured averaged voltage drop (V-I characteristics, Figure S7b) by DC nanovoltmeter (Keithley 2182A) and corresponding oscillation amplitude of voltage difference (dV/dI, Figure S7c) by a lock-in technique. V-I characteristics and dV/dI measurement were measured at the temperature of 2.5 K below T$_C$ of NbSe$_2$. In the direct measurement of dV/dI, clear 4 peaks and minor peaks were shown.

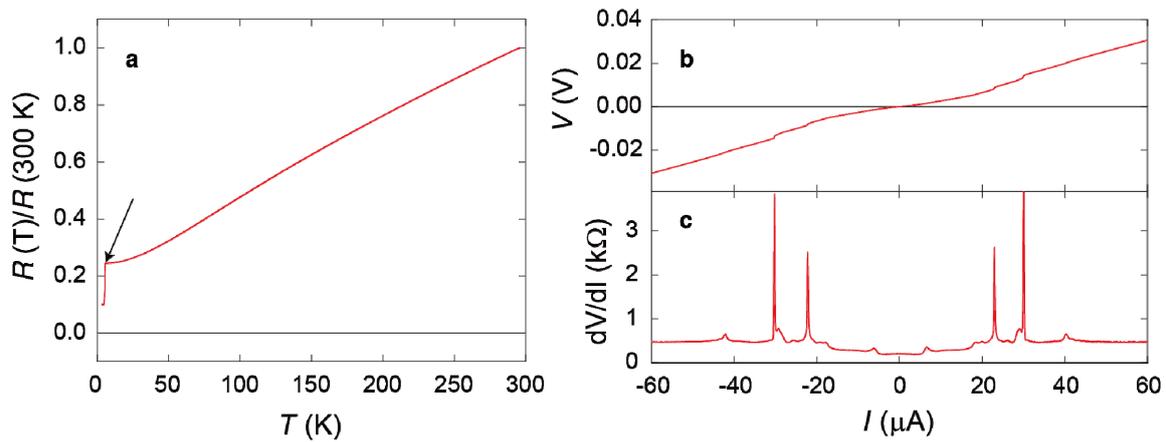

**Figure S8.** Additional transport data for the CrPS$_4$/NbSe$_2$/CrPS$_4$ heterostructure. (a) Temperature dependency of resistance (b) V-I characteristics measured at 2.5 K. (c) Direct dV/dI measurement at 2.5 K. It shows clear 4 peaks with minor peaks.

**Atomic Force Microscope (AFM) image of CrPS$_4$/NbSe$_2$ heterostructure.** AFM (Park systems, NX10) measurement was carried out on CrPS$_4$/NbSe$_2$ heterostructure. The heterostructure shows the clean interface. No wrinkle was observed in the heterostructure. The top-most CrPS$_4$ layer was fresh, showing only minimal polymer residues on it.

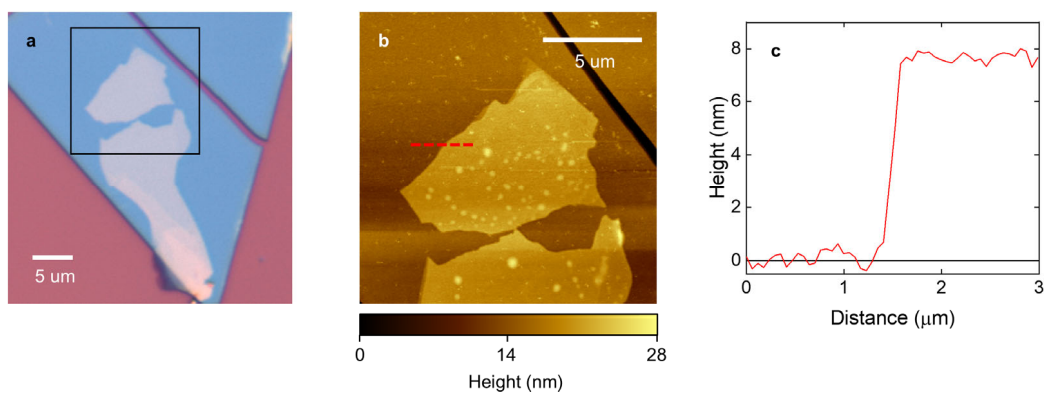

**Figure S9. AFM measurement of CrPS$_4$/NbSe$_2$ heterostructure. (a)** Optical microscope image of the heterostructure. (b) AFM image of selected range (black box in (a)). (c) AFM profile of dashed line in (b).